\documentclass[11pt,a4paper]{article}
\pdfoutput=1
\usepackage{jheppub}

\usepackage{color}
\usepackage{amsmath}
\usepackage{pifont}

\usepackage[utf8]{inputenc}

\usepackage{bbm}
\usepackage{verbatim}   
\usepackage{subfigure}  
\usepackage{acronym}

\usepackage{amsfonts}
\usepackage{amssymb}
\usepackage{mathrsfs}
\usepackage{graphicx}
\usepackage{multirow}
 \usepackage{slashed}

\definecolor{mygreen}{RGB}{29,145,47}
\definecolor{mypurple}{RGB}{164,64,214}
\definecolor{myorange}{RGB}{199,146,32}

\newcommand{\ms}{M_{\odot}}

\newcommand{\meter}{{\, {\rm m}}}

\newcommand{\metre}{{\, {\rm m}}}

\newcommand{\eV}{{\, {\rm eV}}}

\newcommand{\MeV}{{\, {\rm MeV}}}
\newcommand{\GeV}{{\, {\rm GeV}}}

\def\beq{\begin{equation}}
\def\eeq{\end{equation}}
\def\bea{\begin{eqnarray}}
\def\eea{\end{eqnarray}}
\def\bitem{\begin{itemize}}
\def\eitem{\end{itemize}}
\newcommand{\bec}{\begin{center}}
\newcommand{\eec}{\end{center}}
\newcommand{\ba}{\begin{array}}
\newcommand{\ea}{\end{array}}
\def\inv{^{\raise.15ex\hbox{${\scriptscriptstyle -}$}\kern-.05em 1}}
\def\lbar{{\lower.35ex\hbox{$\mathchar'26$}\mkern-10mu\lambda}} %lambda bar

\let\<=\langle
\let\>=\rangle

\let\+=\uparrow

\let\del=\nabla

\begin{document}

\title{Miniclusters in the Axiverse} 
\author[a]{Edward Hardy}
\emailAdd{ehardy@ictp.it}
\affiliation[a]{Abdus Salam International Centre for Theoretical Physics,
Strada Costiera 11, 34151, Trieste, Italy}

\abstract{
If dark matter is an axion-like-particle a significant fraction of the present day relic abundance could be concentrated in compact gravitationally bound miniclusters. We study the minicluster masses compatible with the dark matter relic density constraint. If they form from fluctuations produced by PQ symmetry breaking, minicluster masses up to hundreds of solar masses are possible, although over most of the parameter space they are much lighter. The size of these objects is typically within a few orders of magnitude of an astronomical unit. We also show that miniclusters can form if an axion gets mass from a hidden sector with a first order phase transition that takes a relatively long time to complete. Therefore they can appear in models where PQ symmetry is broken before inflation, compatible with large axion decay constants and string theory UV completions.
}

\maketitle

%%%%%%%%%%%%%%%%%%%%%%%%%%%%%%%%%%%%%%%%%%%%%%%%%

%\input{intro}

\section{Introduction} \label{sec:intro}

Light pseudo-Nambu-Goldstone bosons with an approximate shift symmetry are common in extensions of the Standard Model. Among these, the QCD axion is especially well motivated since it solves the Standard Model strong CP problem. However, more generally it is reasonable to expect multiple QCD-axion-like-particles (which we refer to as axions) and also hidden sector gauge groups with strong dynamics. For example this seems to be the case in compactifications of string theory where axions can come from anti-symmetric tensors wrapping cycles, of which there are typically many \cite{Svrcek:2006yi,Arvanitaki:2009fg,Acharya:2010zx,Cicoli:2012sz}. A feature of such axions is that they can be viable dark matter candidates, with a relic density  depending on their mass, decay constants, and the cosmological history of the universe \cite{Arias:2012az}.

Consequently, it is interesting to study how the phenomenology and observational signatures of axion dark matter at cosmologically late times could differ from those of other dark matter candidates. Possibilities include laboratory searches for axion--photon conversion enhanced by resonance effects, along with other proposals for direct searches some of which rely on the coherence of the axion dark matter field (reviewed in \cite{Graham:2015ouw}). The presence of a background axion population can also lead to observable signals if they decay to photons \cite{Masso:1997ru,Grin:2006aw}.

%One direction of interest is if the axion field thermalises to form a Bose Einstein condensate \cite{Sikivie:2009qn,Erken:2011dz}. If this is such that long range correlations exist, observable differences compared to WIMP models are possible on the scale of galaxies. For example caustics, particular structures in dark matter phase space, can occur giving differences in structure formation \cite{Sikivie:1999jv,Natarajan:2005ut,Sikivie:2010bq} (other possibilities include \cite{RindlerDaller:2011kx,Li:2013nal,2012A37127C,RindlerDaller:2012vj}). The existence of long range order is however disputed \cite{Guth:2014hsa}.

In this note we focus on a feature of axion dark matter that could both modify these signatures and lead to new ones. 
If at early times there are regions of axion over density, these collapse to gravitationally bound objects known as  miniclusters around the time of matter--radiation equality  \cite{Hogan:1988mp,Kolb:1993hw,Kolb:1993zz}. 
The evolution and eventual mass and density of these has been calculated both analytically \cite{Kolb:1994fi} and numerically through N-body simulations \cite{Zurek:2006sy}. Depending on the subsequent dynamics of dark matter halos, there could be observational consequences in current or future microlensing experiments, providing constraints or even a potential discovery. Further, if a significant proportion of dark matter is in miniclusters the signal strength in direct detection experiments will be reduced, and indirect signals modified. In particular, we study the masses and sizes of miniclusters that form, given the observed dark matter relic density and constraints on the scale of inflation from searches for primordial gravitational waves. A detailed analysis of the phenomenology of miniclusters long after their formation is very important, and is left for future work.

Typically the perturbations in axion number density that lead to miniclusters are assumed to be generated by PQ symmetry breaking. In this scenario the mass of the miniclusters is typically only a small fraction of a solar masses, although there are small parts of parameter space where they can be relatively heavy. We also propose an alternative mechanism to generate density perturbations, relying on a strongly coupled hidden sector with a phase transition that happens though bubble nucleation and takes of order a Hubble time to complete. 
This allows miniclusters even if PQ symmetry is broken before inflation, and therefore in theories with large decay constants. Further, axions from string theory are typically in the PQ broken phase during inflation, so such a phase transition can produce miniclusters in this well motivated UV completion.

Finally, it has been suggested that miniclusters, or at least a proportion of the axions within them, could contract further to create Bose stars \cite{Kolb:1993zz,Kolb:1993hw}, with interesting observational signatures. Whether it is possible to construct models with dynamics such that this actually occurs remains unclear, and we do not address this important issue.  Instead, as an intriguing side--note, we analyse the possible range of star masses once the dark matter relic density and star stability constraints are imposed.

Turning to the structure of this paper, in the next Section we summarise some properties of axions and miniclusters. In Section \ref{sec:pqbreaking} we study the parameter space of miniclusters if PQ symmetry is broken after inflation. In Section \ref{sec:phase}, we show how a phase transition could lead to density perturbations and subsequently miniclusters. In Section \ref{sec:stars} we consider the maximum mass of axion stars, and in Section \ref{sec:con} conclude.

%%%%%%%%%%%%%%%%%%%%%%%%%%%%%%%%%%%%%%%%%%%%%%%%%
%%%%%%%%%%%%%%%%%%%%%%%%%%%%%%%%%%%%%%%%%%%%%%%%%

\section{Axions and miniclusters} \label{sec:mini}

For completeness, and to set notation, we begin by reviewing known material on axions and miniclusters. Further details on axions may be found in \cite{Arias:2012az}, and the references therein, while the discussion of miniclusters in this Section follows \cite{Zurek:2006sy}. 

In field theory, axions come from the spontaneous breaking of a global ${\rm U}\left(1\right)$ PQ symmetry at a high scale $f_a$, for example by a field getting a vacuum expectation value (VEV) or strong dynamics in a hidden sector. In contrast, the typical axions from string theory are closed string states and for these there is no energy regime in which the PQ symmetry is linearly realised in a 4-dimensional effective field theory.\footnote{Axions from open strings are also possible and have phenomenology resembling field theory models, since the axion comes from a matter multiplet \cite{Honecker:2013mya,Honecker:2015ela}.} But such axions still have an approximate shift symmetry inherited from a gauge symmetry in the underlying higher dimensional theory. In many compactifications the decay constants of string axions are expected to be close to the string scale, which is often of order $10^{16}\,\GeV$, 
although they could be smaller if the volume of the compactification is exponentially large \cite{Conlon:2006tq}, or in some heterotic models \cite{Buchbinder:2014qca}.

One way for axions to get mass is violation of their shift symmetry by a strongly coupled sector, as is the case for the QCD axion.\footnote{At high temperature, the gauge coupling is small and the QCD axion mass can be understood in terms of instantons.} Using experimental and lattice input the QCD axion mass can be calculated from chiral perturbative theory, and to leading order at zero temperature it is
\beq
m_a^2 = \frac{m_u m_d}{\left(m_u+m_d\right)^2} \frac{m_{\pi}^2 f_{\pi}^2}{f_a^2}  ~,
\eeq
where $m_u$ ($m_d$) is the up (down) quark mass, $m_{\pi}$ is the pion mass, and $f_{\pi}$ the pion decay constant \cite{Weinberg:1977ma}. 
If an axion gets mass from a hidden sector with strong coupling scale $\Lambda$ an analogous dependence is expected
\beq \label{eq:ams}
m_a = \frac{\epsilon^2 \Lambda^2}{f_a} ~,
\eeq
where $\Lambda$ is the hidden sector strong coupling scale. The axion mass can naturally be smaller than $\Lambda^2/f_a$ if the hidden sector PQ breaking vanishes in the limit that a small parameter of the theory goes to zero, corresponding to $\epsilon \lesssim 1$ in Eq.~\eqref{eq:ams}. The QCD axion mass vanishes if one of the bare quark masses is zero, and has $\epsilon \simeq 0.4$ (with $\Lambda_{{\rm QCD}} \simeq 200\,\MeV$). Hidden sectors  must satisfy bounds on the number of additional relativistic degrees of freedom during big bang nucleosynthesis (BBN) and cosmic microwave background observations, and also strong constraints on energy injection into the visible sector after the start of BBN. However, strong coupling scales below an $\MeV$ are not ruled out as long as they do not lead to significant entropy injection, although depending on the number of degrees of freedom a mild temperature difference with the visible sector may be required \cite{Feng:2008mu}.

An axion can also get mass from a sector that does not run into strong coupling, for example an asymptotically free gauge group Higgsed at an energy scale $\Lambda_*$ where it has a coupling constant $g_* \lesssim 1$. Instantons can still lead to an axion mass, but their size is limited to be smaller than $1/\Lambda_*$.\footnote{This is similar to the finite temperature QCD axion mass calculation, in which the gluons have a mass of order $g T$, or equivalently the compactified time dimension cuts off the maximum instanton size \cite{Gross:1980br}.} Consequently, the mass generated is exponentially suppressed, and for a QCD--like hidden sector is parametrically 
\beq
m_a^2 \sim \frac{\Lambda_*^4}{f_a^2} e^{\frac{-8\pi^2}{g_*^2}} ~,
\eeq
with extra dependence on $g_*$ depending on the details of the hidden sector. Similarly, string theory instantons can lead to an axion mass that is exponentially suppressed from other scales.\footnote{Given the expectation that quantum gravity will break all global symmetries such masses are generically expected to be present at some level \cite{Kamionkowski:1992mf}.}

Turning to cosmology, in an expanding universe with Hubble parameter $H$, an axion $a$ evolves as
\beq \label{eq:axem}
\ddot{a} + 3 H \dot{a} - \del^2 a + \partial_a V\left(a\right) =0~,
\eeq
where $V\left(a\right)$ is the temperature dependent axion potential. When $H \gtrsim m_a\left(T\right)$, the axion is constant on scales bigger than the Hubble scale, while on smaller scales the gradient term smooths out fluctuations. This continues until the Hubble parameter drops sufficiently, or the axion mass increases enough, that the potential becomes important and the axion starts oscillating. After a few oscillations it reaches a regime where its comoving number density is conserved, and subsequently behaves as cold dark matter. Depending on the initial field, strings and domain walls can also form, and if present produce axions during their evolution and decay. 

Consequently, the axion relic abundance is affected by the temperature dependence of its mass, which can be parameterised as
\beq \label{eq:sctp}
m_a^2\left(T\right) =  m_a^2 \left(\frac{\Lambda}{T}\right)^n~.
\eeq
If the axion mass comes from a strongly coupled hidden sector, a power law is a reasonable approximation. The  mass is typically suppressed by $ e^{-8\pi^2 c/g^2\left(\mu\right)}$ where $\mu \sim T$, and the gauge coupling runs logarithmically, with $c$ a model dependent constant \cite{Gross:1980br} (close to the strong coupling scale this description will break down). On theoretical grounds it is expected that $f_a \lesssim M_{{\rm pl}}$ \cite{ArkaniHamed:2006dz}, so unless $\epsilon$ in Eq.~\eqref{eq:ams} is very small, or the axion decay constant is close to the Planck scale, the axion mass becomes significant at temperatures slightly above $\Lambda$.

In contrast if the axion mass comes from a Higgsed gauge theory, its mass is constant until temperatures of order the scale at which the gauge theory is broken. Provided this is sufficiently high, the axion starts oscillating when $m_a\left(T=0\right)$ is equal to the Hubble parameter, earlier than in the temperature dependent case. Similarly, if the axions gets its mass from string theory instantons this will be constant until the string scale. The same dynamics occur in a strong coupling model if $\epsilon \ll 1$ in Eq.~\eqref{eq:ams}, and all of these possibilities are reproduced by the $n=0$ limit of Eq.~\eqref{eq:sctp}.

If, in the regime in which it behaves as dark matter, there are order one differences in the number density of axions in different regions of space, miniclusters can form. These are gravitationally bound objects, coming from overdense regions that collapse at the time of matter--radiation equality. Assuming spherical symmetry, the final minicluster density is \cite{Kolb:1994fi}
\beq \label{eq:mcden}
\rho \simeq 140 \Phi^3 \left(\Phi +1 \right) \rho_{c} ~,
\eeq
where $\Phi =  \rho / \left<\rho_c\right>$ is the relative magnitude of the initial density fluctuations, and $\left<\rho_c\right>$ is the average dark matter density at the time of collapse, which is expected to be close to matter--radiation equality. This has also been studied using N-body simulations \cite{Zurek:2006sy}, and it was found that the density of miniclusters is well approximated by the analytic expression Eq.~\eqref{eq:mcden}.

The mass of the miniclusters is determined by the scale of the density fluctuations. The leading expectation is that all the dark matter in a sphere with radius equal to the scale of the density fluctuations, $l$, will collapse. The  mass of the miniclusters $M_m$ is then
\beq \label{eq:mcmassa}
M_m \simeq  5\times 10^{-11} \ms \left(\frac{l}{10^{-18}\GeV^{-1}} \right)^3 \left(\frac{T_0}{\GeV}\right)^3 \left(\frac{g_s}{76} \right) r ~,
\eeq
where $T_0$ is the temperature when the axion begins oscillating, $r$ is the proportion of DM made up by axions, and $g_s$ is the number of relativistic degrees of freedom (which, except for the QCD axion, depends on the details of the hidden sector). It is assumed that there is no entropy injection after the axion begins oscillating.
 In many models of interest, the scale of density fluctuations is parametrically the Hubble length at $T_0$, and in this case
\beq
M_m \simeq 3 \times 10^{-11} \ms \left(\frac{\GeV}{T_0} \right)^3 \left(\frac{76}{g_s} \right)^{1/2} r ~.
\eeq
 The typical radius of miniclusters is fixed by the density Eq.~\eqref{eq:mcden}, and given by
\beq \label{eq:mcra}
R_m \simeq \frac{3 \times 10^{10} \meter}{\Phi \left(1+\Phi\right)^{1/3}} \left(\frac{M_m}{10^{-12} \ms}\right)^{1/3} \left(\frac{0.73 \eV}{T_c} \right)^{4/3}  ~,
\eeq
where $T_c$ is the photon temperature at the time of collapse, expected to be around an eV. For Hubble scale density fluctuations this gives
\beq
R_m \simeq \frac{10^{11} \meter}{\Phi \left(1+\Phi\right)^{1/3}} \left(\frac{\GeV}{T_0}\right)  \left(\frac{76}{g_s} \right)^{1/6} r^{1/3} \left(\frac{0.73 \eV}{T_c} \right)^{4/3}  ~.
\eeq

In the evolution of the universe after matter--radiation equality miniclusters are expected to remain bound, and with a similar mass to when they were formed, although an order 1 fraction of the contained axions could be lost by tidal disruption. The detailed dynamics of miniclusters and the dark matter halo is an important question deserving detailed further study. In particular, it would be interesting to find the present day halo mass function. This would allow a precise determination of the observational signatures or miniclusters. Further, an accurate knowledge of this would allow the impact of miniclusters on axion detection experiments to be quantified. A plausible possibility is that their radius remains similar to Eq.~\eqref{eq:mcra}.

%%%%%%%%%%%%%%%%%%%%%%%%%%%%%%%%%%%%%%%%%%%%%%%%%
%%%%%%%%%%%%%%%%%%%%%%%%%%%%%%%%%%%%%%%%%%%%%%%%%

\section{Miniclusters from PQ symmetry breaking} \label{sec:pqbreaking}

We now apply the material reviewed in the previous section to constrain the parameter space of miniclusters formed from axions. In particular, we extend the discussion of \cite{Zurek:2006sy} to account for the requirement that PQ symmetry is broken after inflation in order that density perturbations are present. Further, in this case (assuming a standard cosmological history) the requirement that the correct axion relic abundance is obtained constrains the viable models (as discussed in, for example, \cite{Arias:2012az}), and therefore we obtain new results on the possible properties of miniclusters.\footnote{This Section is essentially is an extension of the study of miniclusters in the QCD axion case, previous described in \cite{Davidson:2000er}, to more general axion dark matter candidates.} While the computations involved are straightforward, the allowed minicluster properties are strongly constrained.

The observational constraint on isocurvature, along with scale invariance of the power spectrum, prevents significant axion density fluctuations being generated during inflation. Instead, miniclusters typically come from perturbations generated by PQ symmetry breaking. Immediately after PQ breaking the axion field is randomly distributed over the range $a \in f_a \left[0,2\pi\right]$, and if this happens below the energy scale of inflation there will be fluctuations in the axion field over the observable universe.

\begin{figure}
\begin{center} 
 \includegraphics[width=0.7\textwidth]{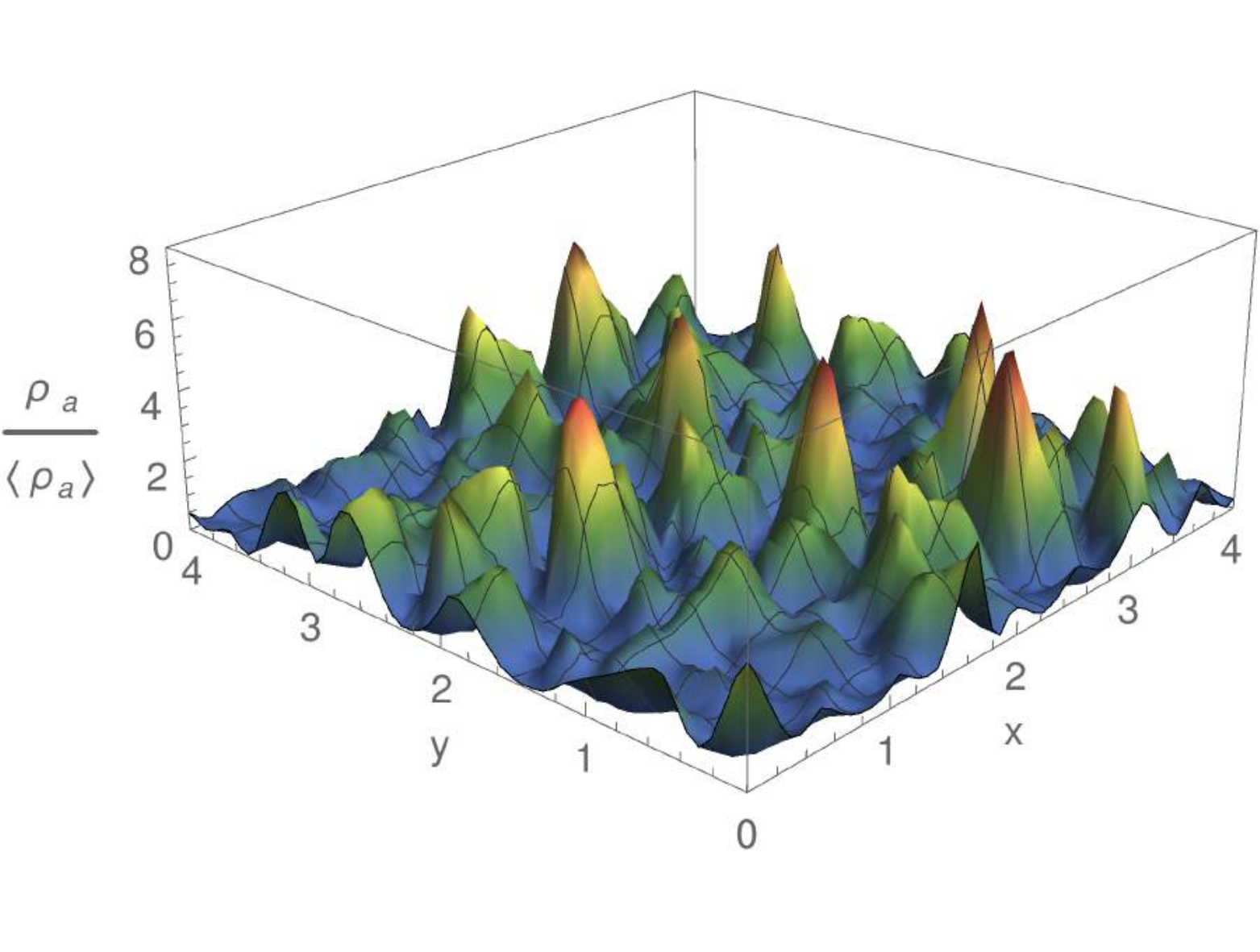}
\caption{The axion energy density, for a typical slice of space, at a time long after the axion has started oscillating but before matter--radiation equality, if PQ symmetry is broken after inflation. Spatial axes are labeled by comoving coordinates, normalised to $1$ at the time when $m_a\left(T\right)=3H$. The random initial field configuration leads to density perturbations on a scale corresponding to the Hubble parameter when the axion starts oscillating, and these will subsequently collapse to miniclusters. Axion strings and domain walls, which are expected to  have a significant effect, are not included in the computation.}
\label{fig:d1}
\end{center} 
\end{figure}

The resulting density fluctuations can be found by numerically integrating the equation of motion, Eq.~\eqref{eq:axem}. This has been studied in \cite{Zurek:2006sy}, starting from an initial condition of white noise long before the axion mass is significant. We repeat the computation, and the axion relic density across a typical slice of space is shown in Fig.~\ref{fig:d1}, at a late time once the axion comoving number density is conserved. Throughout we take the potential to be a cosine, motivated by instantons, however the results obtained are not significantly different for other periodic functions. 

As expected, fluctuations on scales corresponding to those smaller than the Hubble parameter when the axion starts oscillating are smoothed out, while larger density fluctuations remain. The magnitude of the density fluctuations is typically $\rho/ \langle \rho \rangle \sim 8$. This is larger than expected if the axion 
relic density scaled as $\theta_0^2$ where $\theta_0$ is the initial misalignment angle, because of an anharmonic enhancement close to the top of the potential. For $\theta_0\simeq \pi$ the potential  flattens and the 
axion starts oscillating later. As well as enhancing the density fluctuations, this leads to an average axion relic density equivalent to taking $\theta_0 \simeq 2.2$ everywhere in space, rather than $\pi/\sqrt{3}$. There is roughly one fluctuation per Hubble volume, and N-body simulations find that most of the dark matter in a fluctuation collapses \cite{Zurek:2006sy}. Therefore Eq.~\eqref{eq:mcmassa} is a good approximation to the eventual minicluster mass, and an order 1 fraction of dark matter axions are trapped in miniclusters.

So far we have ignored axion strings and domain walls. However, if PQ symmetry is broken after inflation these are generically expected to be present, with a density of roughly one per Hubble patch. Their evolution and decays will produce axions, and alter the dynamics of the rest of the axion field.  Some numerical studies suggest that the presence of strings and domain walls leads to an order 1 increase in the eventual axion relic abundance \cite{Kawasaki:2014sqa}, although these are carried out in parameter ranges very far from the physical values, and there is still considerable uncertainty on the calculation.

The effect of strings and domain walls on minicluster formation is an important issue. To address this requires a full numerical simulation, which we do not attempt in the present work. One possibility is that strings and domain walls produce extra axions that contribute to the smooth axion background density. Assuming such axions make up an order 1 fraction of the dark matter, there will be a proportional reduction in the mass of the miniclusters, and also a decrease in the relative size of fluctuations $\rho_a/\langle\rho_a\rangle$. Alternatively, they might contribute to miniclusters. In this case the mass of the miniclusters will still be constrained by the dark matter mass contained within a Hubble volume when the axion begins oscillating. However, since perturbations are possible on smaller scales without being smoothed out, this scenario could also lead to smaller lighter miniclusters.

\begin{figure}
\begin{center}
 \includegraphics[width=0.47\textwidth]{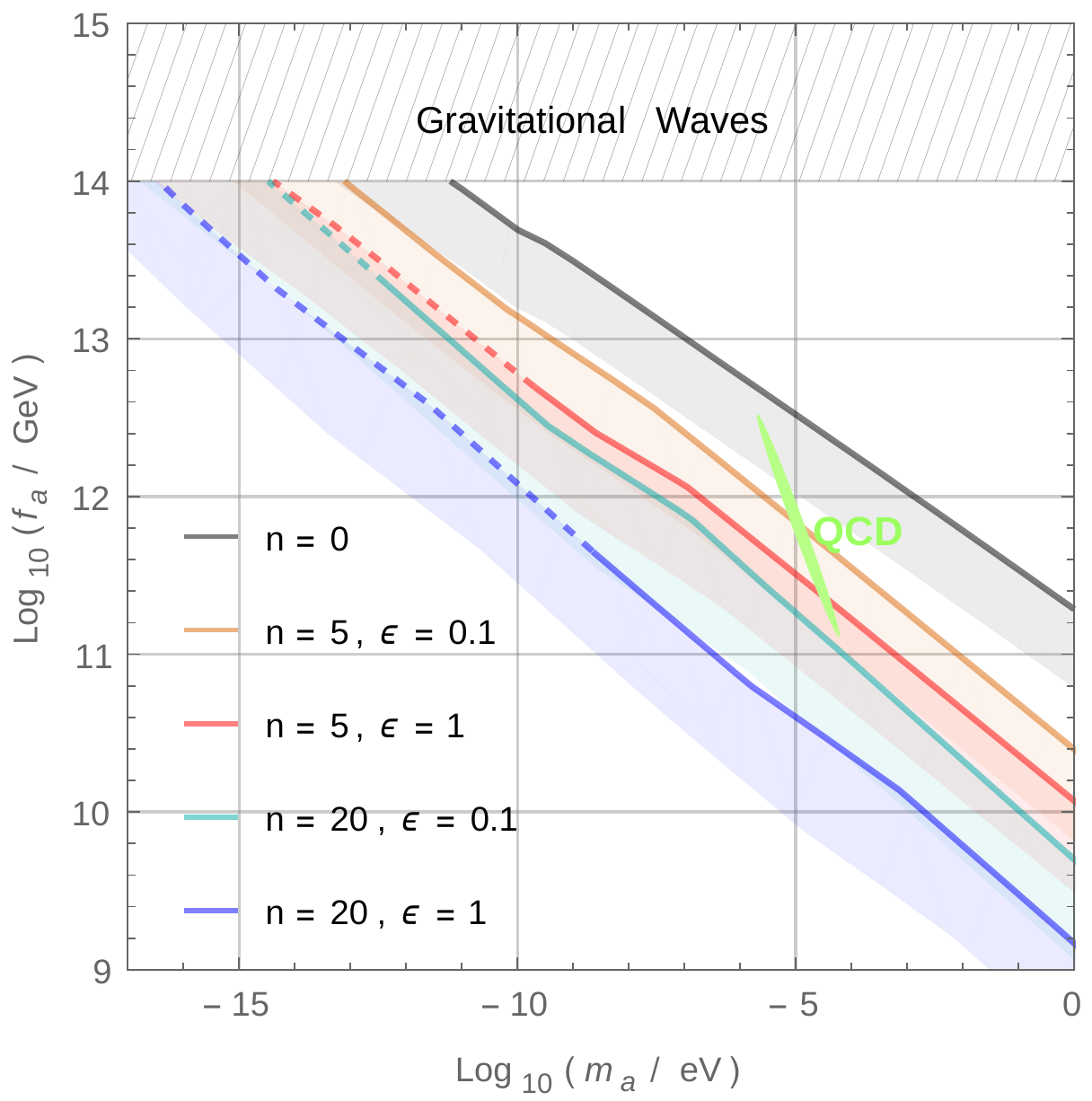}
 \qquad
 \includegraphics[width=0.47\textwidth]{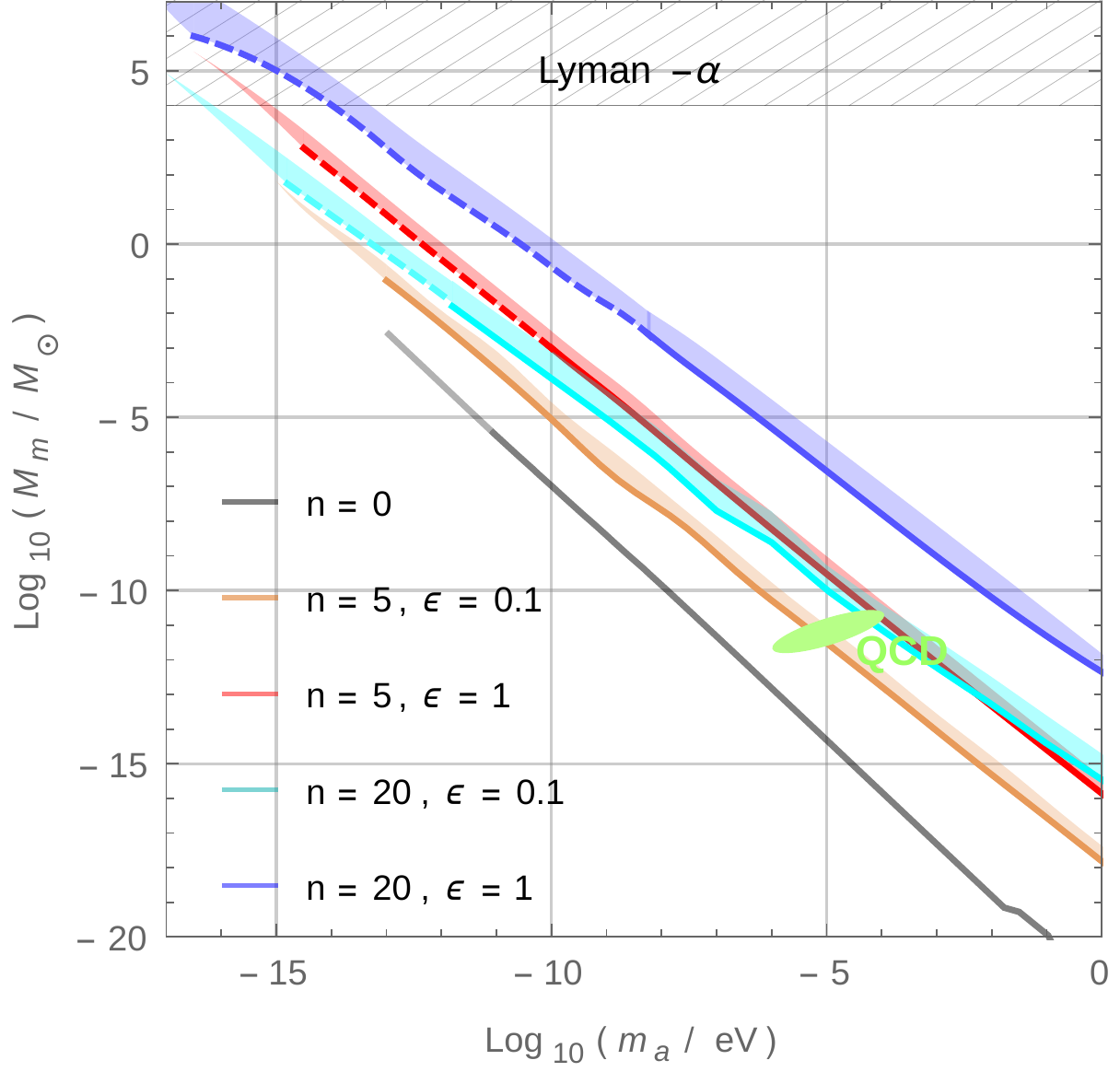}
\caption{{\bf \emph{Left:}} The relationship between axion mass and decay constant for the correct dark matter relic abundance, if PQ symmetry is broken after inflation. This is shown for different axion mass temperature dependences, parameterised by Eq.~\eqref{eq:ams} and Eq.~\eqref{eq:sctp}. Solid lines have a hidden sector strong coupling scale above an $\MeV$, and the axion decay constant is constrained by primordial gravitational wave searches. The part of parameter space corresponding to the QCD axion, with an approximate indication of the uncertainty from the temperature dependence of its mass, is shown in green. The shaded bands indicate the uncertainty arising from strings and domain walls, assuming these could give a relic abundance up to 10 times that from misalignment. {\bf \emph{Right:}} The mass of the largest miniclusters possible in the different axion models. The lines are cut off at axion masses that, for the correct relic abundance, violate the constraint $f_a < 10^{14}~\GeV$ shown in the 
left panel. With the exception of models with a very sharp mass turn on this is the strongest limit on the maximum minicluster 
mass, surpassing that from Lyman--$\alpha$ observations. The QCD axion, shown in green, leads to relatively light miniclusters. Uncertainty from strings and domain walls is again shown shaded.}
\label{fig:1} 
\end{center} 
\end{figure}

The requirement that PQ symmetry is broken after inflation, and that the axion makes up the majority of dark matter, strongly constrains the viable parameter space. Assuming a standard cosmology, for a given model of its mass, the axion relic abundance is completely fixed. Ignoring strings and domain walls, in Fig.~\ref{fig:1} left we plot the relationship between the axion mass and decay constant needed for the correct dark matter relic density, for different models. The lines plotted cover a range of mass temperature dependences from $n=0$ to $n=20$ in Eq.~\eqref{eq:sctp}. Parameter points corresponding to hidden sector strong coupling scales $\Lambda > \MeV$ are plotted as solid lines, and those for $\Lambda < \MeV$ are shown dashed (for all points $\Lambda > 0.01~\MeV$). While the later are not entirely excluded, only some classes of models will evade constraints from BBN. The upper bound on the scale of inflation from primordial gravitational waves,
 $H_I \lesssim 10^{14}~\GeV $, limits the possible axion masses, and the lightest viable axions have a mass of order $10^{-16}~\eV$ and a strong temperature dependence.

The negative gradient of the lines in Fig.~\ref{fig:1} left is because the present day axion relic abundance is proportional to
\beq
\begin{aligned}
\rho\left(T_{now}\right) & \propto  f_{a}^2 m_a\left(0\right)  m_{a}\left(T_{0}\right) \frac{T_n^3}{T_{0}^3} ,\\
\end{aligned}
\eeq
where $T_0$ is the temperature when the axion begins oscillating, and $T_n$ is the temperature today. For models with $n$ order 1 in Eq.~\eqref{eq:sctp}, $T_0 \simeq \Lambda$, and $m_{a}\left(T_{0}\right) \simeq \Lambda^2 / M_{{\rm pl}}$, so that $\rho\left(T_{n}\right) \propto \epsilon m_a^{1/2} f_{a}^{3/2}$ (dropping a dimensionful constant). Meanwhile if the axion mass is constant $\rho\left(T_{n}\right) \propto m_a^{1/2} f_{a}^{2}$. These  relationships roughly fit Fig.~\ref{fig:1} left, although there are differences due to the finite time the axion takes to reach constant comoving number density, along with effects such as the changing number of visible sector relativistic degrees of freedom.

Since the relic density constraint fixes the axion decay constant in terms of its mass, the temperature the axion starts oscillating and the properties of the miniclusters are also fixed. In Fig.~\ref{fig:1} right we plot the mass of the miniclusters, assuming they contain all the dark matter in a Hubble volume when the axion has completed one oscillation (by which time it is in the cold dark matter regime to a good approximation). In Fig.~\ref{fig:1} right, the plotted lines finish at axion masses that require $f_a = 10^{14}~\GeV$. Parameter space to the left of the lines is not accessible in viable models, and the possible minicluster masses are constrained for a particular axion mass. 

The figures also show the uncertainty on the required value of $f_a$ allowing strings and domain walls to give an axion relic abundance that is up to a factor 10 greater than that from misalignment (shaded). The maximum minicluster mass is calculated assuming the axions produced by string and domain wall dynamics have spatial distribution such that they are collapse into miniclusters, although this assumption requires further study.  For fixed axion mass, if strings and domain walls contribute significantly to the the relic abundance, smaller $\Lambda$ and therefore smaller $f_a$ is required. If axions from strings and domain walls contribute to miniclusters, this leads to heavier miniclusters.

Instead of a smooth power law, the axion mass could turn on discontinuously due to a hidden sector phase transition. In Appendix \ref{apa}, it is shown that in this case the viable parameter space with PQ breaking after inflation is very constrained, and does not allow for lighter axions, or heavier miniclusters than in Fig.~\ref{fig:1}. In the next section, we discuss a mechanism allowing miniclusters to form with PQ symmetry broken before inflation in this scenario.

The allowed masses of miniclusters is also constrained by Lyman--$\alpha$ observations. The phase transition adds to the power spectrum (at wavenumber up to the scale at which density fluctuations are smoothed out), and this  must not exceed the cold dark matter contribution at experimentally accessible scales. The limits obtained have been studied in \cite{Afshordi:2003zb,Zurek:2006sy}, where details may be found, and depend on the fraction of dark matter contained in miniclusters. We impose the conservative bound that the minicluster mass must be less than $10^4 \ms$. In all cases except for a very sharp mass turn on, the actual mass of miniclusters is below this value. For the majority of models they are very light compared to a solar mass, and have a radius that is typically less than $10^{13}~\metre$, comparable to the size of the solar system. For a given axion mass, the largest miniclusters occur with a sharp mass turn on since this corresponds to the axion starting to oscillate at the latest time.

The point in model space corresponding to the QCD axion is also shown, with the finite size indicating the uncertainty from the temperature dependence of the mass around a $\GeV$. This leads to an uncertainty on the size of required decay constant, and therefore also on the minicluster mass. Further, the effects of strings and domain walls are not included in the error estimate, and these could plausibly change $f_a$ and the mass of miniclusters by up to an order of magnitude.

Finally, we note that the constraints obtained in this sector could be relaxed if inflation happened at a scale $\lesssim 10^{14}~\GeV$, but the reheating temperature was higher. This would restore PQ symmetry after inflation, even for $f_a \gtrsim 10^{14}$. Allowing larger decay constants would open up the parameter space for light axions, and extend the plotted lines, leading to heavier miniclusters. However, the details of reheating, and whether the universe reaches thermal equilibrium fast enough for the PQ symmetry to be restored, is complex and model dependent \cite{Davidson:2000er}.

%%%%%%%%%%%%%%%%%%%%%%%%%%%%%%%%%%%%%%%%%%%%%%%%%
%%%%%%%%%%%%%%%%%%%%%%%%%%%%%%%%%%%%%%%%%%%%%%%%%

\section{Miniclusters from a phase transition} \label{sec:phase}

If instead PQ symmetry is broken before inflation, the initial axion misalignment is constant over the observable universe. This can be small so larger  decay constants are possible, albeit at the expense of tuning. As mentioned, both large decay constants and PQ symmetry broken before inflation are expected for open string axions, so this is a well motivated scenario. However, if miniclusters are to form a new source of density perturbations is required.

One way this could happen is if there is a first order phase transition in the sector that gives the axion its mass, that occurs though the slow expansion of bubbles. 
It is plausible that the axion mass is dramatically larger inside the bubbles than outside. In particular, we assume it has a discontinuity across the phase transition, and in the high temperature phase it is suppressed sufficiently that $m_a\left(\Lambda+\epsilon\right) \lesssim H\left(\Lambda+ \epsilon\right)$. This is in contrast to the QCD axion, the mass of which is the same over the observable universe at each moment in time because QCD has a crossover.\footnote{It might be possible to make the QCD phase transition first order by introducing a large lepton asymmetry \cite{Schwarz:2009ii}, allowing the QCD axion to form miniclusters through the mechanism in this Section.}

An example of a theory in which the axion mass has these properties is ${\rm SU}\left(N\right)$ Yang-Mills in the large $N$ limit. The zero temperature behaviour of this theory is tractable \cite{Witten:1979kh}, and it has also been studied extensively at finite temperature \cite{Thorn:1980iv} (see \cite{Lucini:2012gg} for a recent review). There are theoretical arguments that in the large $N$ limit, the deconfining phase transition is first order, based on the large difference in number of degrees of freedom across in the different phases \cite{Pepe:2005sz,Poppitz:2012nz}, and this is strongly supported by lattice studies \cite{Lucini:2003zr,Lucini:2005vg}. The axion mass has also been studied  on the lattice. At zero temperature it is found to be non-zero and approximately $m_a = 0.15 \sigma/f_a $ where $\sigma \sim \Lambda^2$ is the string tension \cite{DelDebbio:2002xa}. Meanwhile at finite temperature, the axion 
mass is almost constant up to the critical temperature, when there is a discontinuity and it is very suppressed (and maybe zero) above the phase transition \cite{Lucini:2004yh,DelDebbio:2004vxo}. This behaviour is further supported by theoretical models and holographic arguments \cite{Parnachev:2008fy,Bonati:2013tt,Bigazzi:2015bna}. It is plausible that the required behaviour also occurs in many other models with a first order phase transition. However information is limited by the relatively few theories for which the topological susceptibility has been studied on the lattice (or for which there are controlled theoretical calculations).

\begin{figure}
\begin{center}
 \includegraphics[width=0.47\textwidth]{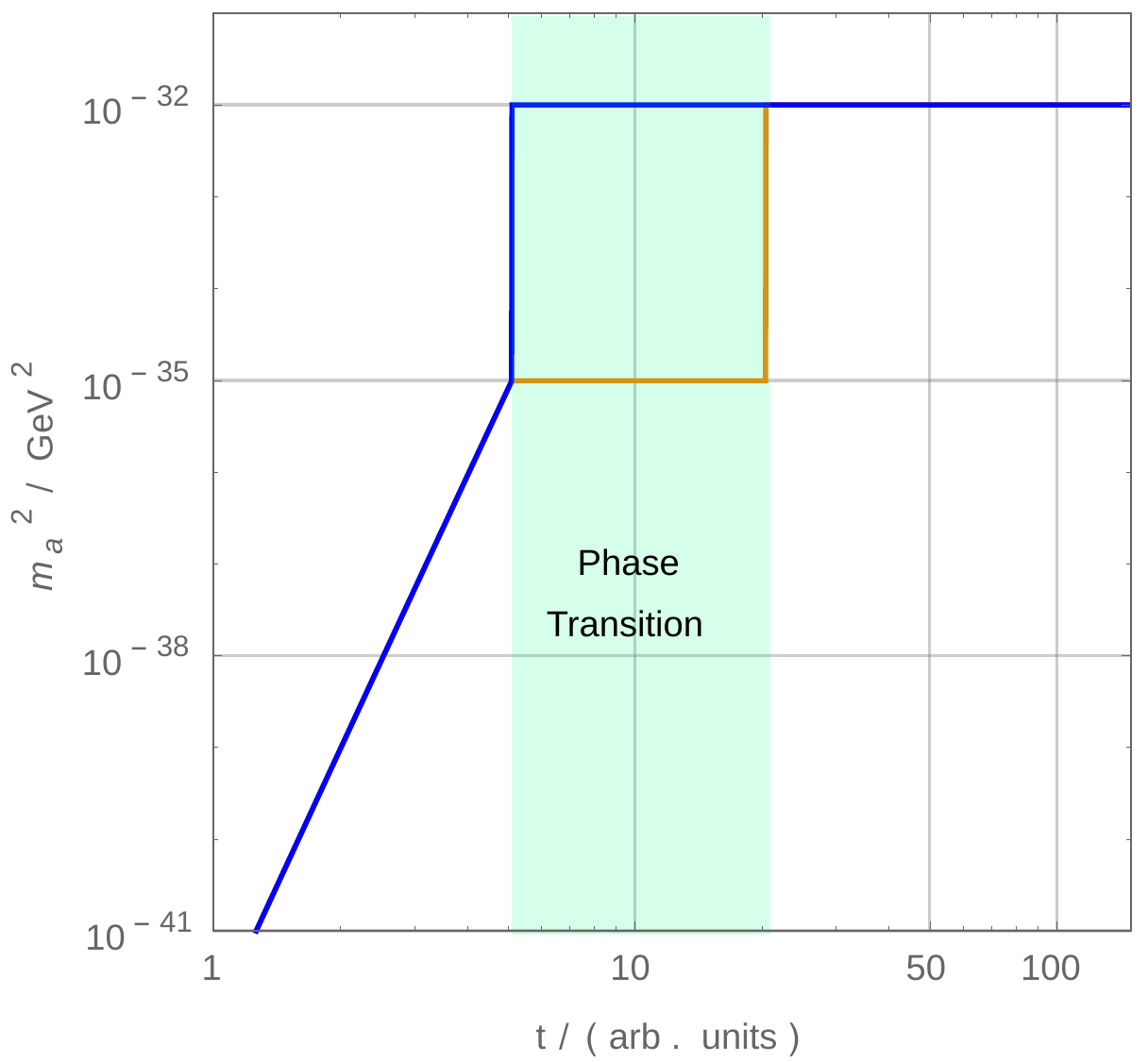}
 \qquad
 \includegraphics[width=0.47\textwidth]{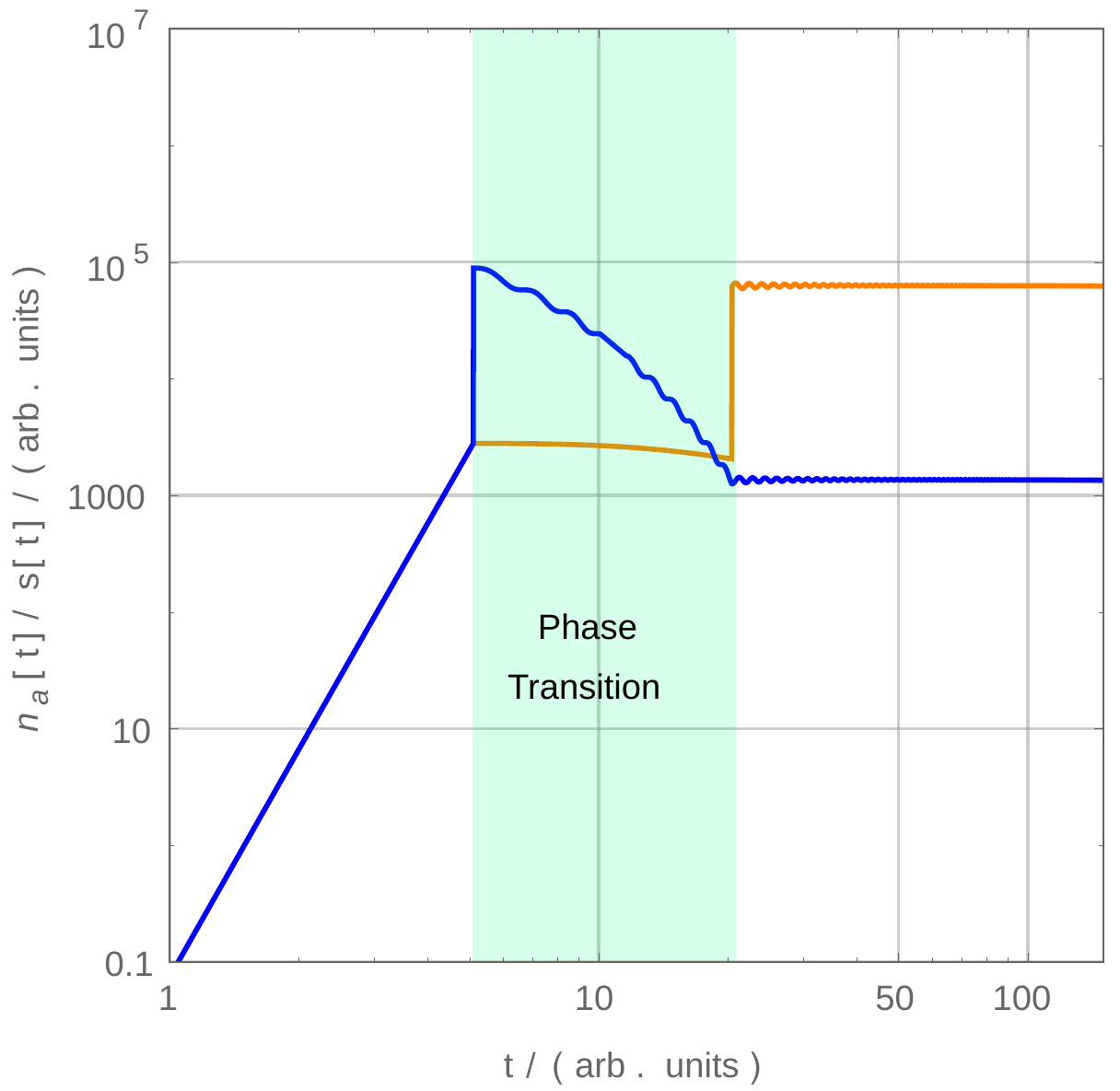}
\caption{{\bf \emph{Left:}} The axion mass as a function of time, in regions of space that transition to the low temperature phase early (blue), and remain in the high temperature phase until late (orange), because of the time bubble expansion takes. It is assumed that there is a discontinuity in the axion mass between the high and low temperature phases, and that the axion mass is constant at its zero temperature value in the low temperature phase. During the phase transition the temperature of the universe is assumed to be constant. {\bf \emph{Right:}} The axion comoving number density as a function of time in parts of space corresponding to the left panel. In regions where the axion begins oscillating early, the number density is depleted by the expansion of the universe during the phase transition. Meanwhile in regions where it gets a large mass late the axion field remains at its initial 
values during most of the phase transition, and reaches a larger number density.}
\label{fig:3}  
\end{center} 
\end{figure}

In such a theory, the axion begins rolling down its potential towards $\theta = 0$ at different times in different parts of the universe, depending when they are engulfed by a bubble. The relic density of axions depends on the time at which it begins oscillating since the subsequent dilution by the expansion of the universe begins once it has reached a solution in which it has a conserved number density. Therefore, regions that start oscillating late will have an dark matter overdensity relative to regions that start oscillating early.\footnote{Another mechanism by which the a first order phase transition could lead to variations in the dark matter density is studied in \cite{Witten:1984rs}.}

The hydrodynamics of the bubbles is complicated and model dependent \cite{Espinosa:2010hh}. However, the qualitative effect on the axion abundance can be captured simply by assuming that the Hubble parameter is constant during the transition. Then the variation in the eventual axion relic density between regions of the universe in which it starts oscillating late relative to regions in which it oscillate early is approximately 
\beq \label{eq:odph}
\frac{\rho_{a,max}}{\rho_{a,min} } \sim \frac{R(t_{max})^3}{R\left(t_{min}\right)^3} \sim e^{3 H\left(t_{max}-t_{min}\right)} ~,
 %\frac{\Omega_{a}}{\Omega_{DM}} \sim \left(\frac{\Lambda}{0.1\,\GeV}\right) \left(\frac{f_a}{10^{16}\,\GeV}\right)^2 \theta_0^2 ~,
\eeq
where $t_{max}$ ($t_{min}$) is the time when the axion starts oscillating in the regions where the axion gets a mass late (early), and $R\left(t\right)$ is the scale factor of the universe at these times. Given a long phase transition,  significant density differences are possible. On scales smaller than the Hubble scale at this time, fluctuations will be smoothed out, so similarly to the PQ breaking case there will typically be one minicluster per Hubble volume, with a typical maximum mass set by the enclosed dark matter.

In Fig.~\ref{fig:3} left, possible time dependences of the axion mass in regions that go through the  phase transition early and late are plotted. A strong power law temperature dependence in the high temperature phase is assumed. In Fig.~\ref{fig:3} right the corresponding axion number density in the two regions are shown (ignoring the spatial gradient contribution in Eq.~\eqref{eq:axem}). In the region where the axion starts oscillating at the start of the phase transition, its number density is significantly diluted, whereas in the other case it begins oscillating only at the end of the transition.

\begin{figure}
\begin{center}
% \includegraphics[width=0.49\textwidth]{dist.pdf}
% \,
 \includegraphics[width=0.7\textwidth]{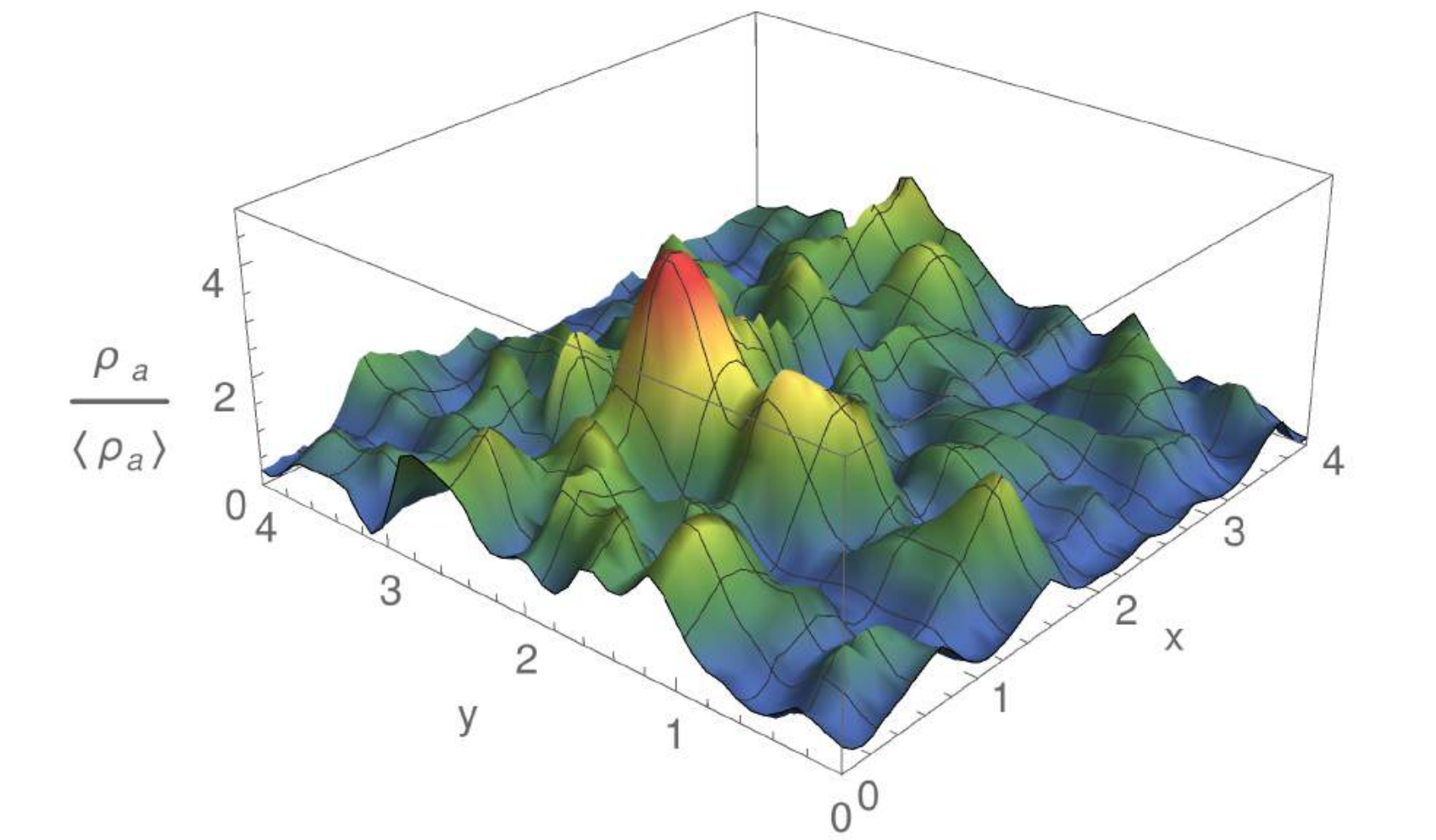}
\caption{%{\bf \emph{Left:}} The distance away from the closest center of a bubble as a function of position in space, for a typical slice of space and a random distribution of bubbles. {\bf \emph{Right:}} 
The axion energy density in a slice of space, after the axion mass has turned on through a first order phase transition, and before matter--radiation equality. The bubbles of the low temperature phase are assumed to expand symmetrically, with a starting density of one per Hubble volume, and fill the entire space in approximately half a Hubble time. Regions that undergo a phase transition later have a higher axion density than those that go through the phase transition early. The spatial coordinates are comoving and normalised to $1$ at the time of the phase transition.}
\label{fig:c} 
\end{center} 
\end{figure}   

For significant density fluctuations to occur, the hidden sector phase transition must take of order a Hubble time to complete. This is a non-trivial constraint on the hidden sector, since in the absence of phase coexistence (described below), phase transitions typically proceed over a timescale $\sim \left(10 H\right)^{-1} \div \left(100 H\right)^{-1}$  \cite{Kosowsky:1992rz}. However, as we now discuss, hidden sectors with the required axion mass behaviour are plausible candidates to have a significantly longer phase transition.

In a first order phase transition as the temperature of the universe decreases the thermal potential develops a second minimum. At the critical temperature $T_c$, this is degenerate with the high temperature minimum in which the universe begins. Subsequently, the bubble nucleation rate $\Gamma$ increases, until it becomes significant at a temperature $T_n$ defined by $\Gamma\left(T_n\right)\sim H^4$. If this condition is never satisfied the phase transition typically fails to successfully complete. In many models the bounce action depends strongly on the temperature of the universe, which in the absence of significant heating of the radiation bath by the phase transition, typically means that phase transitions that successfully complete do so in a small fraction of a Hubble time  \cite{Kosowsky:1992rz,Grojean:2006bp}. 
 \footnote{However, this is not the case for all models, 
for example dynamics that are close to conformal can easily lead to long phase transitions \cite{Konstandin:2011dr}.}

However, as the bubbles of the low temperature phase expand energy is released. If this is sufficient to reheat the universe to the $T_c$ the pressure difference inside and outside the bubble vanishes, known as phase coexistence.  Consequently, the size of the bubbles only increases at a speed parametrically determined by the rate Hubble expansion decreases the temperature of the universe. The thermodynamics of the situation have been studied in detail in \cite{Davis:1999ii,Megevand:2007sv}, and it has also been proposed that a long lasting phase transition could significantly reduce the relic abundance of thermal dark matter candidates  \cite{Wainwright:2009mq}.

Following \cite{Megevand:2007sv}, if $T_n$ is not significantly smaller than $T_c$, the entropy density of the universe during the transition is $s = s_+ \left(1-f\right) +f s_-$, where $s_{+}$ ($s_{-}$) is the entropy density of the high (low) temperature phase, and $f$ is the fraction of the universe in the low temperature phase. Then, since the universe is close to equilibrium, the expansion of the universe during phase coexistence can be calculated
\beq \label{eq:rax}
\frac{R(t_{max})^3}{R\left(t_{min}\right)^3} = \frac{s_+}{s_-} \sim  \frac{s_+}{s_-}\sim \frac{1}{1-\Delta s/s_+}~,
\eeq
where $\Delta s = s_+ - s_-$. From Eq.~\eqref{eq:odph} this is parametrically the magnitude of axion density perturbations that will be generated. The latent heat of the transition $L$ is the difference between the energy densities of the high and low temperature phases at $T_c$ (and is non-vanishing even though the free energies of the high and low temperature phases are equal at $T_c$), and can be related to $\Delta s$ by $\Delta s = L / T_c$.

Consequently, for theories with a large latent heat, the phase transition can take a significant length of time, and lead to order 1 axion density perturbations. In the case that $T_n$ is significantly smaller than $T_c$, the picture is similar, however the latent heat must be larger to reheat the universe up to $T_c$ for phase coexistence. Derivations of the conditions for this to occur and the resulting ratio $R(t_{max})^3/R\left(t_{min}\right)^3$ may be found in \cite{Megevand:2007sv}.

Of course it is desirable to have example theories with the required behaviour. In \cite{Megevand:2007sv}, simple perturbative models were studied, in which the thermal potential can be explicitly calculated. It was found that there are large parts of parameter space where phase coexistence leads to the phase transition taking at least of order a Hubble parameter to complete. Further, in \cite{Wainwright:2009mq} a detailed numerical study of the phase transition of a specific model was performed, finding that the ratio Eq.~\eqref{eq:rax} could be order 1.

Unfortunately it is harder to calculate the full details of a phase transition in a strongly coupled theory, which is the natural setting for the mechanism for minicluster production described here. Many gauge theories have a first order phase transition, for example ${\rm SU}\left(N\right)$ for a wide range of number of fermion flavours \cite{Schwaller:2015tja}, and as discussed are likely to have the required axion mass behaviour. The latent heat of an ${\rm SU}(N)$ gauge theory is expected to scale as $N^2$ in the large $N$ limit, and has also been studied on the lattice, from which $L \simeq 0.3 N^2 T_c^4$ \cite{Lucini:2005vg}. Evaluating the absolute value of the entropy of the high temperature phase is harder, since this requires lattice results to be regularised (whereas the energy difference across the phase transition does not have this issue \cite{Lucini:2005vg}). It may be very roughly estimated from the free theory value 
\beq
s_+ \sim \frac{2 \pi^2}{45} g T_c^3 \sim \frac{4 \pi^2}{45} \left(N^2-1\right) T_c^3~,
\eeq
where $g$ is the effective number of relativistic degrees of freedom, and we have assumed that the hidden sector dominates the energy density of the universe. 

Although these estimates are far from precise, they suggest that it is plausible that ratio $\Delta s/ s_+$ could be significant allowing for phase coexistence in ${\rm SU}\left(N\right)$ theories. A full computation of the phase transition for such theories would also require an understanding of the process of bubble nucleation and expansion and release of latent heat, along with the temperature $T_n$ at which this becomes significant. These depend on the full effective potential, and obtaining reliable precise results is likely to be hard. However, given the large latent heat, along with results from simpler models, for our present purposes we assume that a phase transition timescale of roughly a Hubble time is reasonable.

\begin{figure}
\begin{center}
 \includegraphics[width=0.47\textwidth]{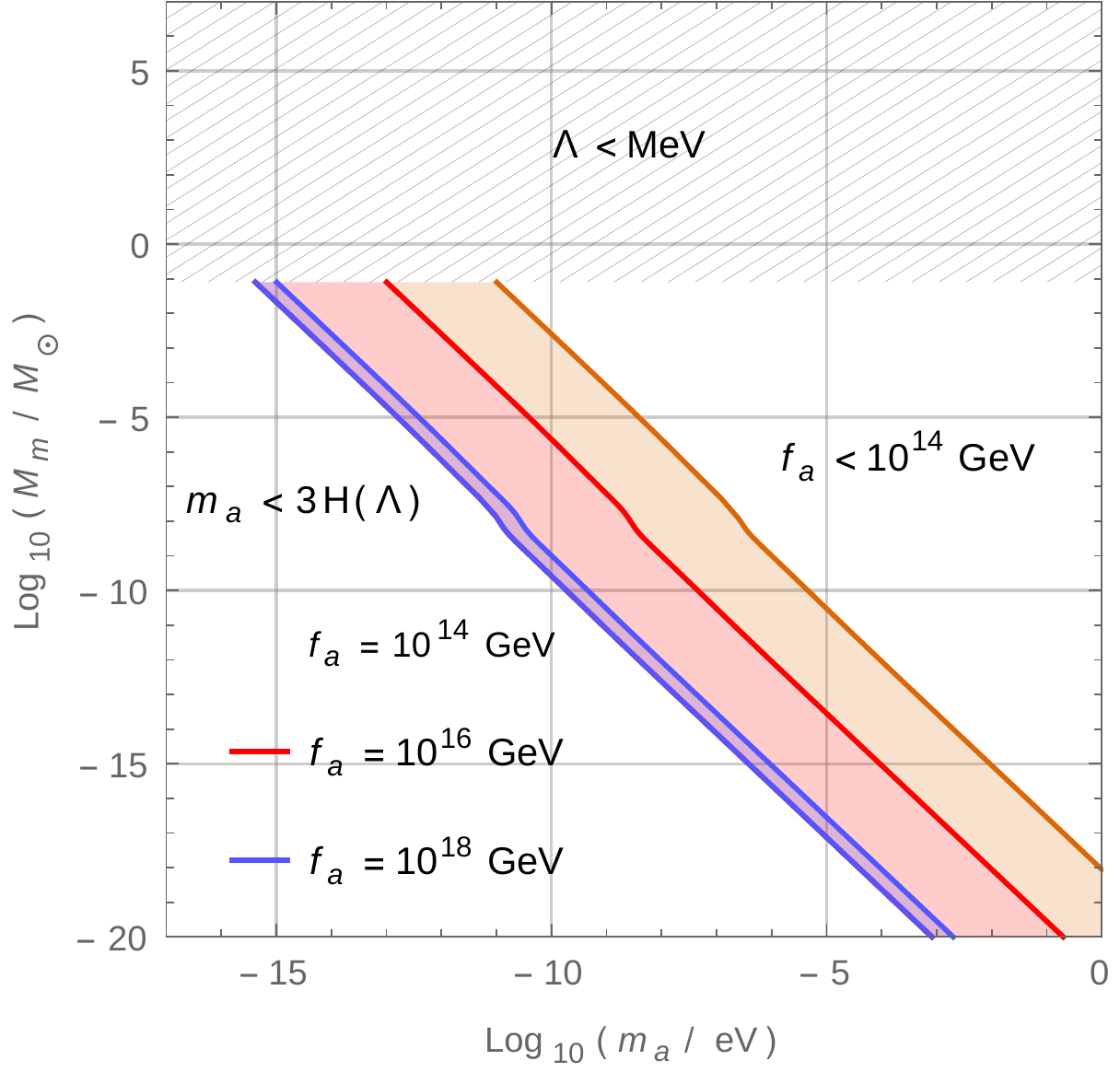}
\caption{The maximum minicluster mass for different values of the axion decay constant, if they are formed by the dynamics of bubbles during a phase transition. The shaded regions overlap and models with all values of $f_a$ can reach the left most blue line. BBN constrains the strong coupling scale $\Lambda$ to be above an MeV. Models with  $f_a \lesssim 10^{14}~\GeV$ can populate the white region in the upper right of the plot, but the lower left white region does not lead to miniclusters, since models here have $m_a < 3 H\left(\Lambda\right)$.}
\label{fig:6}  
\end{center} 
\end{figure}

To study the miniclusters produced by such a hidden sector, the axion equations of motion can be numerically solved in a toy model of a phase transition. Seed points of bubbles are put randomly in space, with a density of 1 per Hubble volume (and we assume no new bubbles form during the transition). These are assumed to expand spherically symmetrically, with constant speed such that the whole universe is in the low temperature phase after  half a Hubble time.
%In Fig.~\ref{fig:c} left we plot the distance from the closest bubble as a function of the point in space for a typical slice of space. We also assume a sharp axion mass turn on, so this is the time delay before the axion starts oscillating in each part of space. 
In Fig.~\ref{fig:c} the resulting late time axion density is plotted for a slice of space.  The fluctuations are close to the expectation from Eq.~\eqref{eq:odph}, and have a size approximately fixed by the Hubble scale during the phase transition. We stress that the dynamics of realistic phase transitions are much more complicated than assumed here, but the miniclusters produced are likely to be similar.

The  properties of the minclusters are constrained by the observed dark matter relic density, similarly to the PQ broken case in Section \ref{sec:pqbreaking}. Since a first order phase transition is needed, the hidden sector strong coupling scale must be above an $\MeV$ to avoid constraints from BBN. The parameter $\epsilon$ in the axion mass Eq.~\eqref{eq:ams}  satisfies $\epsilon \lesssim 1$, since PQ symmetry breaking occurs at the scale $\Lambda$. For miniclusters to be produced, the axion must start oscillating immediately in the low temperature phase (rather than at a later time, long after the phase transition has finished). The condition for this is $m_a > 3 H\left(\Lambda\right)$, giving a lower bound $\epsilon \gtrsim \sqrt{f_a/M_{\rm pl}}$.

Therefore, for a particular $f_a$, there is a range of allowed $\epsilon$ and $\Lambda$, and for the parameters of interest the correct relic abundance can always be obtained by choosing $\theta_0$. The masses of the miniclusters obtained are plotted in Fig.~\ref{fig:6}, for example values of the decay constant, over the allowed parameter range. The maximum mass of the miniclusters is approximately $0.1 \ms$ from Eq.~\eqref{eq:mcmassa}  because of the constraint on $\Lambda$ (with an uncertainty of a factor a few depending on the number of hidden sector degrees of freedom). To the left of the region corresponding to $f_a =10^{18}~\GeV$ miniclusters do not form, while in white region to the right of the plot minicluster can form in models with smaller decay constants than are shown.

%%%%%%%%%%%%%%%%%%%%%%%%%%%%%%%%%%%%%%%%%%%%%%%%%
%%%%%%%%%%%%%%%%%%%%%%%%%%%%%%%%%%%%%%%%%%%%%%%%%

\section{Axion stars}\label{sec:stars}

There are stable Bose star configurations of axions, which are supported by pressure \cite{Chavanis:2011zi,Chavanis:2011zm,Barranco:2008tr,Davidson:2016uok}, and
it has been suggested that these could be produced by further contraction of miniclusters \cite{Kolb:1993zz,Kolb:1993hw}.  If they do exist, axion stars lead to distinctive observational signatures  for example through gravitational lensing \cite{Kolb:1995bu}, or even 
gravitational wave emission in collisions \cite{Giudice:2016zpa}. Collisions between axion stars and neutron or white dwarf stars could also lead to photon signals \cite{Iwazaki:1999my,Barranco:2012ur}. In our present work we do not investigate whether this is dynamically possible, and instead study the allowed axion star masses in realistic models, assuming miniclusters are required for their formation. Similarly to Section \ref{sec:mini}, the physical properties of axion stars have been studied in the literature \cite{Kolb:1993zz,Kolb:1993hw}. Our present contribution is the simple observation that requiring PQ symmetry breaking after inflation, combined with obtaining the correct dark matter relic abundance, constrains the possible axion star masses as a function of the axion mass. Consequently, an observational hint of miniclusters in a particular range would strongly motivate direct searches for axions in the corresponding mass range, and vice versa.

\begin{figure}
\begin{center} 
 \includegraphics[width=0.47\textwidth]{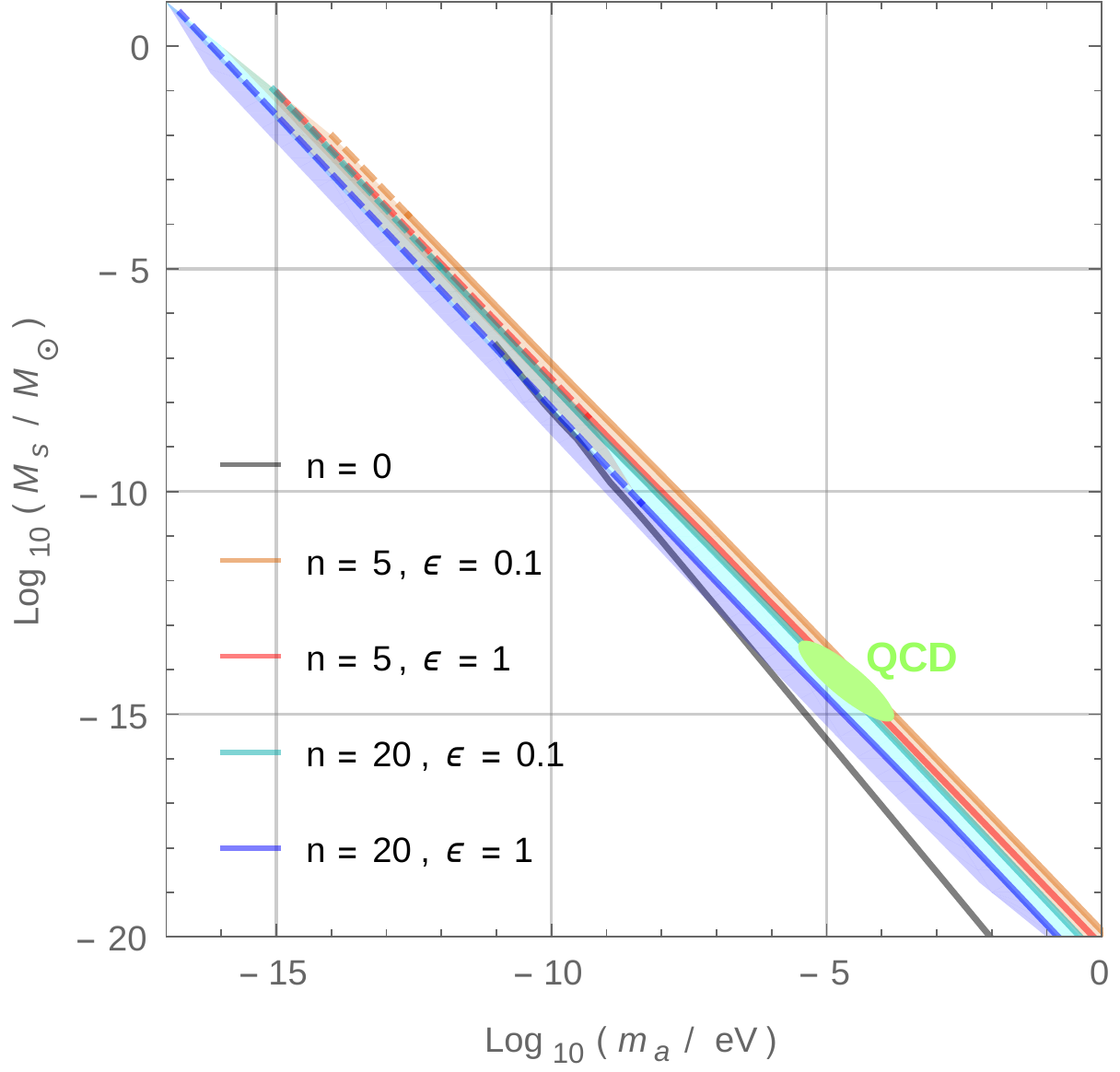}
\caption{The maximum axion star mass possible in models where PQ symmetry is broken after inflation, given the stability condition Eq.~\eqref{eq:starc}. The lines are cut off by the gravitational wave constraint, $f_a \lesssim 10^{14}~\GeV$, in combination with the requirement that the correct dark matter relic density is obtained. Solid lines correspond to parts of parameter space with a hidden sector strong coupling scale above an MeV. Comparing with Fig.~\ref{fig:1} right, the stability constraint is stronger the imposing that the star mass is smaller than the mass of typical miniclusters.  The uncertainty from the unknown contribution of strings and domain walls to the relic abundance is shaded, and models corresponding to the QCD axion are shown in green.}
\label{fig:star}  
\end{center} 
\end{figure}  

We assume that the axion has an attractive quartic interaction, and that this is not unnaturally small. A repulsive quartic would improve star stability, but could make the dynamics of star formation even harder to realise. The stability bound on an axion Bose star is well known \cite{Chavanis:2011zi,Chavanis:2011zm}
\beq \label{eq:starc}	
M \lesssim \frac{f_{a} M_{{\rm pl}} }{m_a}~,
\eeq
and above this the star is expected to either fragment into smaller objects or collapse to a black hole. 

If the miniclusters are produced by PQ symmetry breaking the possible star masses are very strongly constrained. For a given axion mass temperature dependence, the decay constant is fixed and therefore we can also fix the maximum axion star mass, which is plotted in Fig.~\ref{fig:star}. As in Fig.~\ref{fig:1} right, the lines are cut off when dark matter constraint requires the axion decay constant is greater than $10^{14}\,\GeV$, incompatible with gravitational wave searches. Over all of the parameter space, only relatively light axion stars are possible, and stability is a stronger constraint than imposing that the star mass is less than the mass of miniclusters (shown in Fig.~\ref{fig:1} right).

Alternatively, miniclusters could be formed from the phase transition as in Section \ref{sec:phase}. In this case heavier axion stars can be stable, since much larger axion decay constants are possible, since the misalignment angle can be tuned small. Consequently, the stability condition Eq.~\eqref{eq:starc} does not constrain the interesting parameter space at all for $f_a \gtrsim 10^{16}~\GeV$. Instead axion stars as large as the miniclusters in Fig.~\ref{fig:6} are possible if  appropriate dynamics allows them to form. However, this still does not include any parameter points exceeding a solar mass.

%%%%%%%%%%%%%%%%%%%%%%%%%%%%%%%%%%%%%%%%%%%%%%%%%
%%%%%%%%%%%%%%%%%%%%%%%%%%%%%%%%%%%%%%%%%%%%%%%%%

\section{Conclusion}\label{sec:con}

In this short note we have studied the masses of miniclusters that can form in models of axions, once the requirement that the correct dark matter relic density is imposed, and accounting for the different possible temperature dependences of the axion mass. While the computations involved are straightforward, the constraints on the possible masses of miniclusters are strong. Further, if either an axion or astrophysical objects with the properties of miniclusters are detected, the relations we give would strongly motivate complementary searches in specific parts of parameter space.

As well as the standard scenario in which miniclusters come from perturbations generated during PQ symmetry breaking, we have proposed an alternative scenario in which miniclusters form because of the dynamics of the gauge sector that gives the axion its mass. It is also possible there are other mechanisms by which suitable density perturbations could be generated. Also, we have only considered a standard cosmological history. It would be interesting to extend our study to include non-thermal cosmological histories, which could alter the accessible parameter space. 

The lack of a proper numerical simulation of the contribution of axions from strings and domain walls in the case of PQ breaking after inflation is a major deficiency of our present work. These are expected to alter the axion relic abundance by at least an order 1 factor, and it is unclear how they will affect the density perturbations that contribute to the formation of miniclusters. This is especially important for interpreting experimental searches for the QCD axion that target such models. However, proper simulation of the axion string and domain wall system is numerically difficult and obtaining reliable results is challenging.

A second area for future work is the dynamics of the miniclusters at late times, since we have only evolved the axion field to a time shortly after they start behaving as dark matter. For the PQ breaking after inflation scenario, the gravitational collapse of miniclusters at matter-radiation equality has been examined using an N-body simulation in \cite{Zurek:2006sy}. It would be worthwhile to carry out a similar study for the bubble phase transition case.  The dynamics of this could lead to differences in the spatial distribution of high density regions, and the properties of miniclusters. Further, the miniclusters formed at matter-radiation equality must be evolved forward to the present day. This is essential to allow for a proper understanding of the potential exclusion and discovery in microlensing searches, and would allow a reliable computation of the proportion of dark matter in miniclusters, which is important for axion searches.

Finally, it would be interesting to study if there are any other dark matter candidates that could have significant density perturbations arising from the dynamics of hidden sector phase transitions, analogously to those in Section \ref{sec:phase}. These would collapse to similar minicluster like objects, and could lead to interesting phenomenology.

\section*{Acknowledgements}
I am grateful to Robert Lasenby, David Marsh, and Giovanni Villadoro for very useful discussions.

\appendix

\section{Miniclusters with a sharp axion mass turn on} \label{apa}

In this appendix we discuss the miniclusters formed from PQ breaking after inflation, in the case of an instantaneous axion mass turn on. 
%which was not considered in Section \ref{sec:pqbreaking}
Since $\theta_0$ is fixed, obtaining the correct relic density in this case requires
\beq \label{eq:aa1}
f_a = \frac{7\times 10^{-9}~\GeV}{\epsilon^6} \left(\frac{\eV}{m_a} \right)~,
\eeq
where we have taken the number of relativistic degrees of freedom $= 76$ at the time the axion starts oscillating, and the axion mass is defined as in Eq.~\eqref{eq:ams}.

As discussed in Section \ref{sec:phase}, an instantaneous axion mass turn on is typically associated with a first order phase transition. Therefore, we restrict the strong coupling scale $\Lambda > \MeV$ to avoid constraints from BBN. This could be evaded by assuming a significant temperature difference between the visible and hidden sectors, which would allow lighter axions. Consequently, we require
\beq
\begin{aligned}
\epsilon & < 1000 \sqrt{m_a f_a}/\GeV \\
& < 0.04~,
\end{aligned}
\eeq
using Eq.~\eqref{eq:aa1}. Demanding that the axion begins oscillating immediately after turning on (which is assumed in  Eq.~\eqref{eq:aa1}), places a lower bound on $\epsilon$. This is that $m_a > 3 H\left(\Lambda \right)$, and therefore
\beq
\epsilon > 7 \times 10^{-10} \left(f_a/\GeV \right)^{1/2} ~.
\eeq
Otherwise the axion effectively has no temperature dependence for the purposes of the relic density calculation, a possibility which is already accounted for in Fig.~\ref{fig:1} by the $n=0$ line.

\begin{figure}
\begin{center}
 \includegraphics[width=0.47\textwidth]{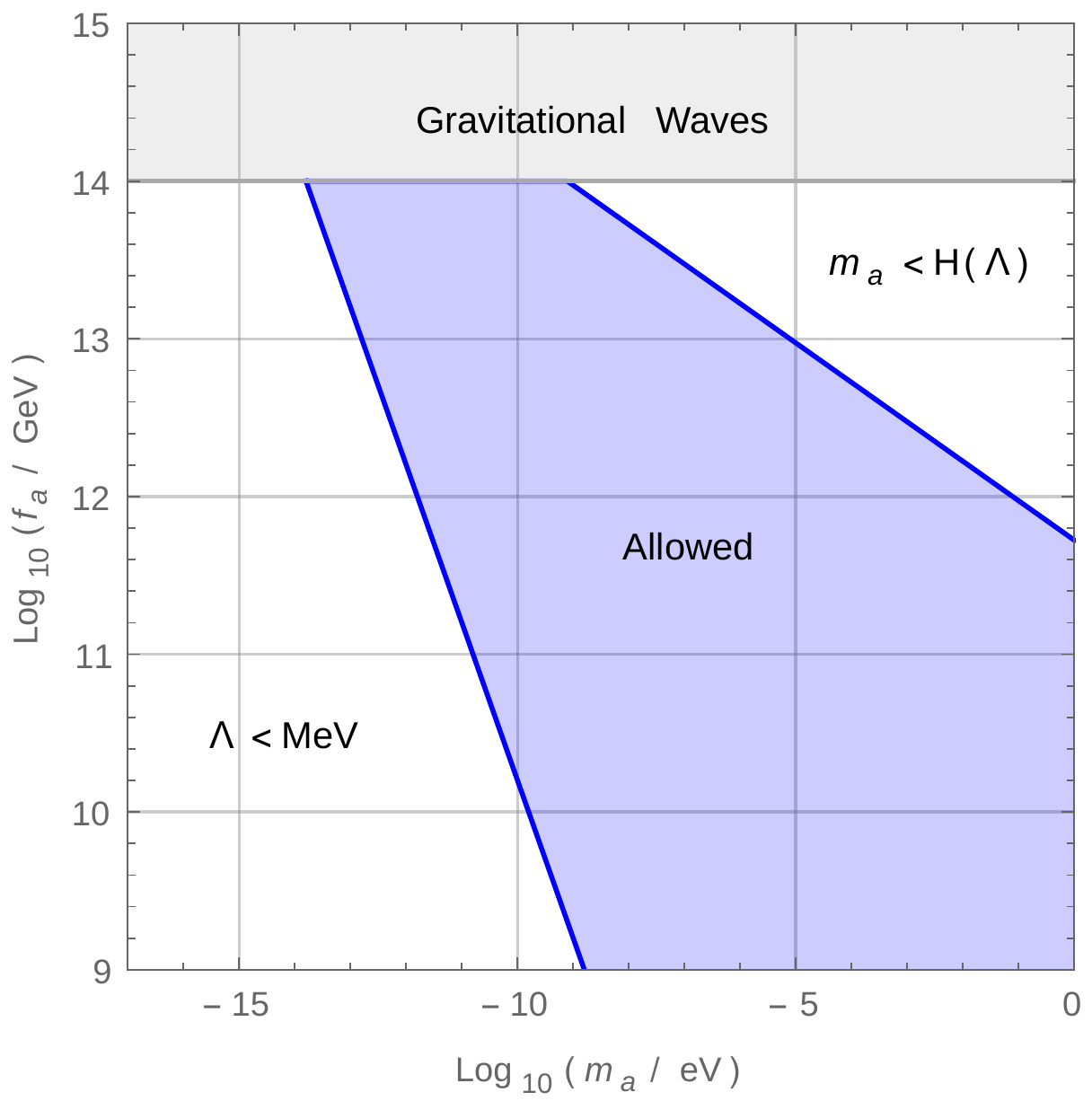}
 \qquad
 \includegraphics[width=0.47\textwidth]{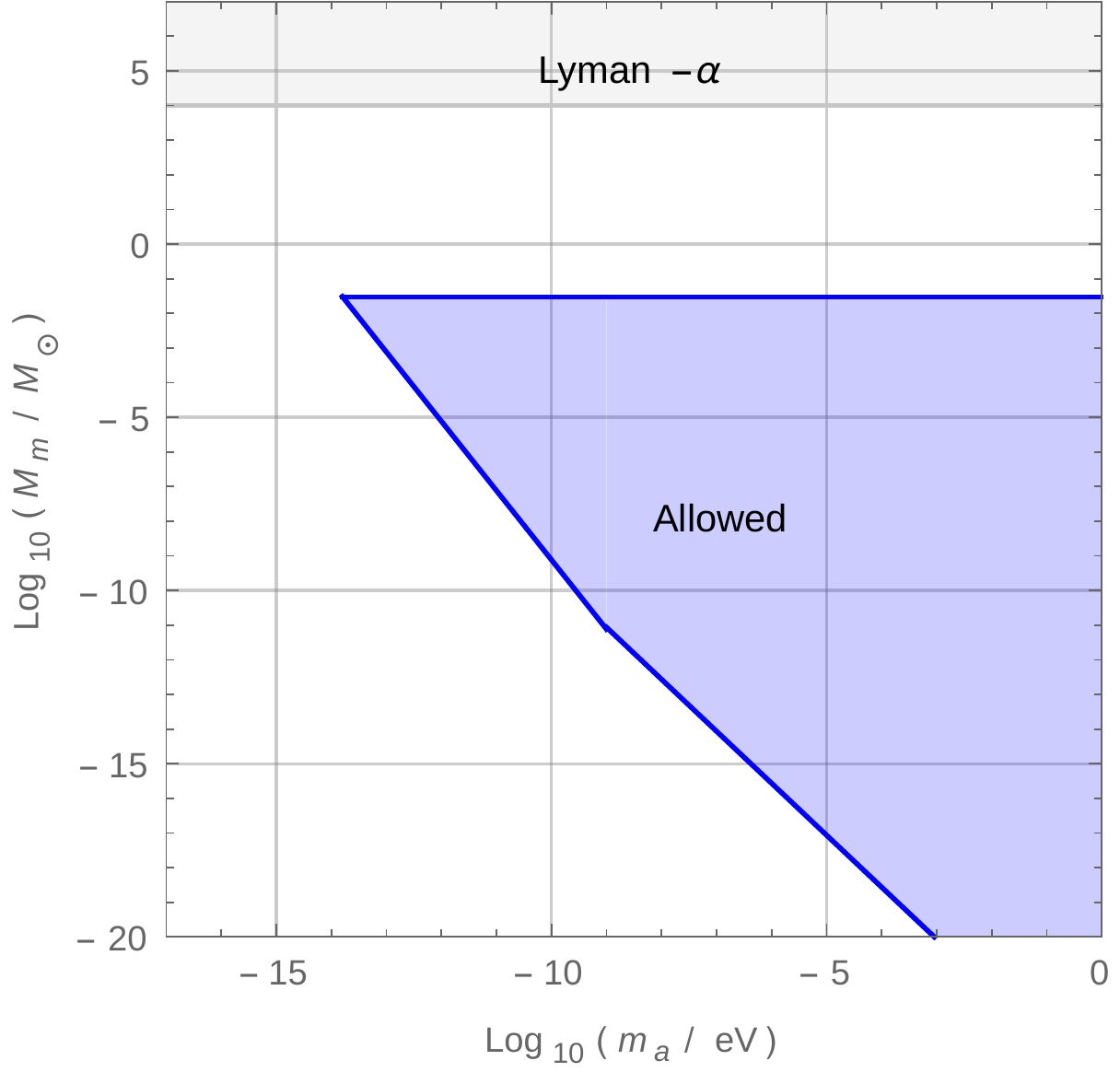}
\caption{{\bf \emph{Left:}} The parameter space that can reproduce the correct dark matter relic abundance if PQ symmetry is broken before inflation, and the axion mass turns on instantaneous. We impose that this occurs at a scale $\Lambda$ above an MeV, to avoid constraints from BBN since such transitions are typically first order. {\bf \emph{Right:}} The masses of miniclusters formed in such models. The constraint that $\Lambda > \MeV$  means that only relatively light miniclusters are possible.}
\label{fig:ap} 
\end{center} 
\end{figure}

Therefore, there is an allowed region in the $f_a$ against $m_a$ plane in which the correct relic density is possible, which is plotted in Fig.~\ref{fig:ap} left. Similarly, the possible masses of miniclusters in this scenario can be obtained since these are fixed by $\Lambda$ through Eq.~\eqref{eq:mcmassa}, plotted in Fig.~\ref{fig:ap} right. Because of the restriction on the minimum size of $\Lambda$ the maximum minicluster masses possible are relatively small $\sim 0.1 \ms$, and axions with mass less than $10^{-14}~\eV$ are not possible.

Finally we note that the values of $\epsilon$ that are viable are small (and, further, the allowed range is not large in many cases). Consequently, in realistic models there may be hidden sector states that are light relative to $\Lambda$, analogous to the pions in QCD, which could be cosmologically dangerous.

\bibliographystyle{JHEP}
\bibliography{reference}

\end{document}